\documentstyle[12pt]{article}
\begin{document}
\begin{titlepage}

\title{Dispersion Relations and Effective Field Theory}

\author{John F. Donoghue\\ [5mm]
Department of Physics and Astronomy\\
University of Massachusetts
Amherst, MA ~01003 ~U.S.A.}

\date{ }
\maketitle

\begin{abstract}
The techniques of dispersion relations match very well with those of
effective field theory. I describe the techniques for using dispersion
relations effectively, and give some pedagogical examples to
illustrate the range of applications. 
\end{abstract}
{\vfill UMHEP-426, hep-ph/9607351  Lecture 
presented at the International School on
Effective Field Theory, Almunecar,  Spain, June 1995}
\end{titlepage}

Dispersive techniques were common in the 1960's, but have been less
popular since then due to the triumph of field theory. Why should we
bother to resurrect them at this time? Besides being a useful aspect of
the toolkit for any theorist, I feel that they have a special utility
in effective field theory. More particulars will be given below, but
briefly stated they are useful in effective theories because they can
accurately describe the low energy propagation of the light degrees of
freedom as well as, and sometimes better than, perturbative Feynman
diagrams. This can lead to new and/or better types of calculation as
well as increasing our insight into the workings of effective field
theory. 
 
In this lecture, I briefly explore the methods, successes and
limitations of combining dispersion relations with effective field
theory. The goal is to first demonstrate that one can reproduce the
low energy content of Feynman diagram calculations in an effective
theory by simply using the lowest order vertices as the input to a
dispersion relation. Then we address what is needed to improve on this
lowest order calculation and how to do the matching of the dispersion
relation with the effective field theory method. I will describe
various examples to illustrate the method, and finally summarize the
advantages and disadvantages of this approach.[1]

\section{Dispersion relations - general}

Scattering amplitudes and vertex functions will in general contain both real 
and imaginary parts.[2]  The imaginary portions are due to the
propagation of on-shell intermediate states.  Causality 
implies certain properties for the 
analytic structure of the amplitudes that allows us to relate the real and 
imaginary parts.  Such dispersion relations have the general form

\begin{equation}
Re f(s) = {1 \over \pi} P \int_{0}^{\infty} {ds' \over s' -
s}Im  f(s')
\end{equation}

\noindent With the identity

\begin{equation}
{1 \over x-x_0 - i \epsilon} = P {1 \over x-x_0} + i \pi \delta (x-x_0)
\end{equation}

\noindent one can write the full amplitude as an integral over its 
imaginary part,

\begin{equation}
f(s) = {1 \over \pi} \int_{0}^{\infty} ds' {Im f(s') \over s' - s - i 
\epsilon}
\end{equation}

Notice that the dispersive integral involves all $s'$.  In order to know 
$f(s)$ at small $s$, we need to know $Im f(s')$ also at large $s'$.  
We will see that subtractions can lessen the dependence on large 
$s'$, but the integral still runs over all $s'$.  We in general 
need to know the properties of on-shell intermediate states.  

Given a dispersion relation, one may also write a ``subtracted'' relation 
for $(f (s) - f(0))/s$, i.e.,

\begin{equation}
{f(s) - f(0) \over s} = {1 \over \pi} \int {ds' \over s' - s 
- i \epsilon} Im \left[ {f(s') - f(0) \over s'} \right]
\end{equation}

\noindent which, since $Im f(0) = 0$, is equivalent to

\begin{equation}
f(s) = f(0) + {s \over \pi} \int {ds' \over s'} {Im f(s') 
\over s' - s - i \epsilon}
\end{equation}

\noindent Subtractions may be needed 
if $f(z) \neq 0$ as $\mid z \mid \rightarrow \infty$, as a 
good behavior at infinity is required for the derivation of the 
dispersion relation.  However, even if subtractions are not required, it may 
still be desirable 
to perform them.  This is especially true in effective field theories,
where we are interested primarily in the low energy quantum effects,
while we do not know how to calculate the higher energy physics.
Generally the input to the dispersion relation, $Im f(s)$,
 is not well known at high
energy.  The subtracted dispersion integral weights lower energies more
heavily and lessens the influence of the high energy region.  
The previous ignorance of the high energy effects of $Im f(s)$
is  reduced to a single number, the subtraction constant.  We will see
that these subtraction constants are equivalent to the parameters in
the effective lagrangians. Further
subtractions  may be performed, with the introduction of further
subtraction constants. 

\section{Calculating with Dispersion Relations}

The pion form factor obeys dispersion relations.  An unsubtracted form is

\begin{equation}
f_{\pi} (q^2) = {1 \over \pi} \int^{\infty}_{4 m^2_{\pi}} ds' {Im 
f_{\pi} (s') \over s' - q^2 - i \epsilon}
\end{equation}

\noindent while with one subtraction the form is

\begin{equation}
f_{\pi} (q^2) = 1 + {q^2 \over \pi^2} \int^{\infty}_{4m^2_{\pi}} {ds' 
\over s'} {Im f_{\pi} (s') \over s' - q^2 - i 
\epsilon}
\end{equation}

\noindent Here the subtraction constant has been fixed to unity by the 
normalization of the form factor.  A twice subtracted form is

\begin{equation}
f_{\pi} (q^2) = 1 + cq^2 + {q^4 \over \pi} \int^{\infty}_{4m^2_{\pi}} {ds'
\over s'^2} {Im f_{\pi} (s') \over s' - q^2 - i 
\epsilon}
\end{equation}

\noindent where $c$ is presently unknown.
  
The key step in a complete dispersion relation calculation is in
specifying the input into the dispersion integral, i.e. the imaginary
part of the function. In this case, the imaginary part is known
experimentally up to about $1.5 GeV$. The good news is that the
physics of the imaginary part is relatively simple. It is well known
that the largest effect in the pion form factor is the rho
intermediate state. This of course has lead to the technique of 
vector dominance. The simplicity of intermediate states in dispersion
relations can help in their implementation. However before turning to
phenomenology, let us see how this technique can be reconciled with a 
Feynman diagram approach.

All Feynman 
diagrams have the same analytic structure as the amplitude that they 
contribute to. They can therefore be rewritten as 
dispersion relations, most often with subtractions.  Therefore the content of 
chiral loops can equally well be specified as a particular choice 
for $Im f(s')$ in a dispersion integral. Let us see how this occurs
for the pion form factor.

In chiral perturbation theory, a one loop calculation involves all 
the machinery of field theory including a full set of diagrams, 
regularization of divergent integrals, and renormalization of the parameters
in the Lagrangian. This is a well functioning machine and beautifully 
extracts the predictions of the theory. However much of this effort
is irrelevant for the real dynamical content of an effective theory.
The effective theory is only able to predict the low energy behavior of 
the light degrees of freedom of the theory. All of the high energy effects
associated with divergences, regularization and renormalization are 
irrelevant to the final answer. One needs to carry out the full 
field-theoretic procedure because otherwise
the Feynman diagram machine breaks down and gives wrong answers, but
the high energy portions of the calculation 
do not express the real physics of the effective theory. 

In the case of the pion electromagnetic 
form factor, at lowest order $O(E^2)$, one predicts simply that $f_{\pi} 
(q^2) = 1$. The analysis to $O(E^4)$ is described in detail in Ref [3].
The two important ingredients are the low energy constant $L_9$ 
and the effect of loops. The tree level contribution 
involves the parameter $L_9$, which occurs in the general
chiral Lagrangian at order $E^4$ [4], plus other constants i.e.,

\begin{equation}
f^{(tree)}_{\pi} (q^2) = 1 + {2L_9 \over F^2_{\pi}} q^2 +{8m^2_{\pi}
 \over F^2_{\pi}} (2L_4 + L_5)
\end{equation}

\noindent Of the loop diagrams, Fig. 1b, c, that of Fig. 1b has no 
$q^2$ dependence, contributing only a constant. Since 
we know that the pion form factor is absolutely normalized 
to unity at $q^2 = 0$, we know that all corrections which are 
independent of $q^2$ must be canceled by 
the wavefunction renormalization constant along with constant terms in 
Eq. 9 and the diagram of Fig. 1c. 
Thus diagram 1b and the renormalization constant 
are dynamically irrelevant although in  this method of calculation they 
enforce the constraint on the normalization of the formfactor.

However, Fig. 1c is more interesting because it also contains important
dynamical information of the propagation of the two pion state, including
the imaginary part of the amplitude due to on-shell intermediate states,
and the result involves a nontrivial function of $q^2$,

\begin{eqnarray}
\Delta f^{(1c)}_{\pi} (q^2) = {1 \over 16 \pi^2 F^2_{\pi}} \left\{ \left( 
m^2_{\pi} - {1 \over 6} q^2 \right) \left[ {2 \over d-4} + \gamma 
- 1 - ln 4 \pi + ln 
{m^2_{\pi} \over \mu^2} \right] \right.\\ \nonumber
\left. + {1 \over 6} \left( q^2 - 4 m^2_{\pi} \right) H (q^2) - {1 \over
18}  q^2 \right\}
\end{eqnarray}

\noindent with

\begin{eqnarray}
H(q^2) &=& 2 + \beta \left[ ln \left( {1-\beta \over 1 + \beta} \right) + i \pi 
\theta (q^2 - 4m^2_{\pi}) \right] \\ \nonumber
\beta &=& \sqrt{1-{4m^2_{\pi} \over q^2}}
\end{eqnarray}

\noindent Multiplying by the wavefunction renormalization constant
and defining the renormalized value of $L_9$

\begin{equation}
L^r_9 = L_9 - {1 \over 192 \pi^2} \left[ {2 \over d-4} - ln 4 \pi + \gamma 
-1 \right]
\end{equation}

\noindent we get the final result

\begin{equation}
f_{\pi} (q^2) = 1 + {2L^r_9 \over F^2_{\pi}} q^2 + {1 \over 96 \pi^2
F^2_{\pi}} \left[ (q^2 - 4m^2_{\pi}) H(q^2) - q^2 ln {m^2_{\pi} \over 
\mu^2} - {q^2 \over 3} \right]
\end{equation}

All of the dynamical content of the effective theory 
was in the vertices and 
propagation of the two pion state in Fig. 1c. This represents low 
energy-long range propagation that the effective theory is capable of
predicting.

Now let us  get this same physics in a dispersion relation.[5] 
The key feature is the choice of what to use for the imaginary part of
the amplitude.
The one loop diagram, Fig. 1c, involves the $\pi \pi$ I = 1 
scattering amplitude, and the tree level $\pi \pi \rightarrow \gamma$ vertex, 
so that

\begin{equation}
2(p_1 - p_2)_{\mu} Im f_{\pi} (s) = \int {d^3 p_1' d^3 p_2' 
\over (2 \pi)^6 2E_1' 2E_2'} (2 \pi)^4 \delta^4 (s - p_1' - 
p_2') \\ \nonumber 
\langle \pi \pi \mid T \mid \pi \pi \rangle \langle
\pi \pi \mid  J_{\mu} \mid 0 \rangle
\end{equation}

\noindent If we choose the forms for these amplitudes which is predicted 
at lowest order in chiral symmetry, we obtain

\begin{equation}
Im f_{\pi} (s) = {1 \over 96 \pi F^2_{\pi}} {(s - 4m^2_{\pi})^{3 \over 2} 
\over \sqrt{s}} \theta (s - 4 m^2_{\pi})
\end{equation}

\noindent We use this in 
the twice subtracted form, Eq. 8, and the dispersion 
integral can be exactly done using

\begin{equation}
\int_{4m^2}^\infty  {ds \over s^2} \left( 1 - {4m^2 \over s} 
\right)^{1 \over 2} \left( {a + b s 
\over s - 
q^2 - i \epsilon} \right) = {(a + b q^2) \over q^4}H (q^2) - {a \over 
6m^2q^2}
\end{equation}

\noindent to give

\begin{equation}
f_{\pi} (q^2) = 1 + c q^2 + {1 \over 96 \pi^2 F^2_{\pi}} \left[ ( q^2 - 
4m^2_{\pi}) H (q^2) + {2 \over 3} q^2 \right]
\end{equation}

\noindent Comparing this with the chiral calculation, Eq. 7, leads to the 
identification of the subtraction constant

\begin{equation}
c = {2 L^{(r)}_9(\mu ) \over F^2_{\pi}} - {1 \over 96 \pi^2 F^2_{\pi}} \left( ln 
{m^2_{\pi} \over \mu^2} + 1 \right)
\end{equation}

We see that we can reproduce the content of the Feynman diagram
approach simply by using the lowest order vertices and propagators in
the dispersion integral. 
When it is phrased this way, it is clear that 
the content of chiral loop diagrams such as Fig. 1c and the content of a 
dispersive integral are similar.  The chiral calculations uses a predicted 
approximation to $Im f(s')$, while a properly performed dispersion 
integral uses the real world data for $Im f(s')$. We also see that the
chiral parameters $(L_i)$ play a similar role to the subtraction constants
in dispersion relations.

The logic of effective theories is even clearer with dispersion relations 
than with Feynman diagrams, although both are adequate tools. When 
giving talks on chiral perturbation theory, one routinely encounters 
individuals who think that the use of effective Lagrangians in loop 
diagrams is not allowed. Their confusion generally centers on the fact that
effective Lagrangians only predict the low energy behavior, yet loops
involve an integration over all energies, which they mistakenly
interpet as a prohibition on performing loop calculations. 
A dispersive analysis, such as 
is given above, is presumably more acceptable to this way of thinking since,
when twice subtracted, it is only sensitive to the chiral amplitude
at low energies. The equivalence of the dispersive treatment and the 
Feynman diagrams can then be used to demonstrate that the output of 
chiral loops is also the result of the low energy amplitude.

\section{ Relative Strengths} 

The above example does not display the full power of either 
chiral symmetry or dispersion relations. Both can bring new information
to the calculation. Let us discuss these special features before we 
attempt to combine the best of the two methods.

The special feature of the chiral symmetry is that it provides relations
between amplitudes with different numbers of pions. Most symmetries,
such as isospin, relate amplitudes of states that belong to the same
multiplets. For a dynamically broken symmetry with Goldstone bosons, the 
symmetry transformation does not relate states within a multiplet, but
rather it relates states with greater or fewer zero energy Goldstone bosons.
Corrections to this limit can be
given in an 
expansion in the energy.  There exist various reduced matrix elements 
which are not predicted by the symmetry and which therefore must be 
measured. These are the parameters in the chiral Lagrangian.
The outputs of chiral perturbation theory are relations between amplitudes, 
order by order in the expansion in E, $m_q$.  At any given order, these 
relations form low energy theorems of QCD. In the above example, chiral
symmetry is more powerful than dispersion relations when it comes to the
subtraction constant, because an extended chiral analysis can also tell
us that the same constant $L_9$ appears in other reactions. For example,
$L_9$ also contributes to radiative pion decay, $\pi \to e \nu \gamma$,
and can be independently measured in that process.[4]

On the other hand, dispersion relations can ultimately do a better job 
on the low energy dynamical effects, as captured in the dispersive
integral. The one loop chiral analysis is equivalent to the lowest
order vertices and propagators, while a dispersive treatment can use
instead the full answer given by Nature. In this particular example,
there is in fact quite a substantial difference, because there is a
resonance in the I=1 channel of $\pi\pi$ interactions. In Fig. 2, I 
display the integrand of the dispersion relation as a function of energy
$E=\sqrt{s}$. Specifically, for the
twice subtracted relation  
 
\begin{equation}
f_{\pi} (q^2) = 1 + cq^2 + {q^4 \over \pi} \int^{\infty}_{2m_{\pi}} {dE} 
{N(E)}{E^2 \over E^2 - q^2 - i 
\epsilon}
\end{equation}

\noindent with

\begin{equation}
N(E) = {(2E) Im f(E) \over E^6}
\end{equation}

\noindent Fig 2 plots $N(E)$, whose integral determines the low $q^2$
behavior of the dispersive integral. The dashed line is the lowest order 
chiral amplitude that is the integrand discussed in the previous 
section, equivalent to the one loop answer. In contrast, the solid line 
is the full answer from a fit to the experimental imaginary part.
We see that the results agree at very low energy, but that at moderate
energies the resonance
is much larger than the lowest order chiral result. This is not a 
problem of principle because it describes 
physics that would appear in the chiral
expansion at higher order. However it does indicate that the
dispersive analysis can be used to obtain a more accurate answer than
can the one loop chiral analysis.

\section{Matching Conditions}

We have seen that chiral symmetry can provide a more extensive analysis
of the subtraction constants, while dispersion relations are capable of
yielding more accurate information on the intermediate states. This suggests
that it may sometimes be advantageous to combine the best of both techniques.
To do this we need to be able to tie the two formalisms together in the
most accurate way.

In dispersion relations involving subtraction constants we need a precise 
identification of them.  Chiral perturbation theory provides these
constants.  The key is to reformulate chiral calculations as dispersion 
relations, order by order.  An 
important point is that the matching is different at order $E^2$ [6,7]
 and at order 
$E^4$ [5,8].

At order $E^2$ one needs to reproduce only the tree level chiral results, 
which do not involve imaginary parts.  Thus we only need to ensure that the 
normalization at low energy is correct.  The dispersion integral will then 
produce new effects at order $E^4$ which are equivalent to the prediction of 
the low energy constants at order $E^4$, i.e., of the $L_i$.  This procedure 
will be more sensitive to high energy effects because one will be using a 
dispersion integral with at most one subtraction.

At order $E^4$ one knows more about the low energy structure so one can 
use a dispersion relation with an extra subtraction.  The low energy 
constants $L_i$ are no longer predicted, but are inputs to fix the subtraction 
constants [The dispersion integral then produces new effects at order $E^6$ 
and higher].  To match at this order one must reproduce the one loop chiral 
calculation.  Therefore the inputs to the dispersive integral must involve the 
lowest order vertices, and will only have free propagations of the 
intermediate state, i.e., the same inputs that go into the Feynman diagram 
calculation.

To actually carry this out, we need three steps. First, the chiral calculation 
needs to be carried out to the given order. Then we reformulate the problem
as a dispersion relation, requiring that the dispersion relation give
the same result when treated to the same order. This fixes the subtraction
constants in terms of chiral parameters. Finally we need to use a 
representation of the full imaginary part which is compatible with the
low energy chiral constraints. This procedure will exactly reproduce the
chiral calculation to the order that it is valid, yet add more physics at
higher orders in the energy expansion. 

The example of the pion formfactor that was used above illustrates the 
matching technique at $O(E^4)$. We determined the subtraction constant 
in a way which was accurate to $O(E^4)$ by a comparison with a direct 
chiral calculation, and then to complete the calculation we feed the 
experimental imaginary part, given in Fig 2, into the dispersion integral.
The use of the experimental $Im f_{\pi} 
(s)$ then generates the full $f_{\pi} (q^2)$ at all $q^2$.  
In principle, the only inaccuracy in this 
calculation is that we have given the 
subtraction constant c by an expression which is exact only to order $E^4$.  
There can be corrections to this by extra factors of $m^2_{\pi}$ or 
$m^2_{\pi} ln m^2_{\pi}$.

Let us also briefly consider the same quantity matched at $O(E^2)$, using 
Eq. 7.  Now the only matching is the simple constraint $f_{\pi} (0) = 1$, 
and the effect of the dispersive integral starts at $q^2$.  This leads to a 
prediction of the low energy constant

\begin{equation}
2L^r_9(\mu) + {1 \over 96 \pi^2} 
\left( ln {m^2_{\pi} \over \mu^2} + 1 \right) = 
F^2_{\pi} \int^{\infty}_{4 m^2_{\pi}} {ds' \over s'^2} Im 
f_{\pi} (s')
\end{equation}

\noindent Note that the lowest order form of $Im f_{\pi} (s)$ cannot be 
inserted in the once subtracted dispersion integral, as the result diverges.  
The lowest order form for $Im f_{\pi} (s)$ is not valid at high energies, but 
the twice subtracted integral used above was not sensitive to this.
The use of the real data for $Im f_{\pi}(s')$ leads to a successful prediction
of $L^r_9$ in terms of the mass of the rho meson.

As a side point, let me note that the dispersive treatment also gives us 
the answer to a previously puzzling feature in the application of 
vector dominance to the prediction of the chiral parameters.[9,10] 
The renormalized
chiral parameters are scale dependent and it was never clear at what scale
they were supposed to equal the prediction of vector dominance. A commonly 
used but ad-hoc answer is that the predictions were valid at a scale equal
to the mass of the rho meson[10]. The dispersive treatment shows the correct 
result. Rather than predicting the constant $L_9^r(\mu)$ itself, one is 
predicting the physical scale-invariant combination given by the dispersive
sum rule. This has no scale ambiguities, and is directly physically relevant.

Once one has provided an accurate matching of the two techniques, one can use
the dispersive integral to extend the calculation beyond the range of the
chiral result.
Thus in the best of all worlds (full data on $Im f(s)$, 
many related reactions) 
the two techniques form a powerful combination which allows rigorous 
results at all energies.  Chiral perturbation theory provides the subtraction 
constants from symmetry relations and dispersion relations allows the 
extrapolation to higher energy.

\section{Example:  The Weinberg sum rules and some relatives}

The simplest amplitudes are two point functions, and within QCD the 
simplest of these are the particular combination of vector and axial vector 
currents.

\begin{equation}
\pi^{\mu \nu}_V (q^2)-\pi^{\mu \nu}_A (q^2) \equiv i \int d^{4} x 
e^{iq \cdot x} \langle 0 \mid T \left[ V^{\mu} (x) V^{\nu} (0)-A^{\mu} (x) 
A^{\nu} (0) \right] \mid 0 \rangle
\end{equation}

\noindent This combination is analytic in the complex $q^2$ plane, except 
for a pole at $q^2 = m^2_{\pi}$ and a cut for $q^2 > 4 m^2_{\pi}$.  The 
vector current is conserved.  The axial current is conserved in the $m_q 
\rightarrow 0$ limit, but with a Goldstone boson.  If we define scalar 
function by

\begin{eqnarray}
\pi_V^{\mu \nu} (q^2) &=& (q^{\mu} q^{\nu} - g^{\mu \nu} q^2) \pi_V 
(q^2) \\ \nonumber
\pi^{\mu \nu}_A (q^2) &=& (q^{\mu} q^{\nu} - g^{\mu \nu} q^2) \pi_A 
(q^2) - q^{\mu} q^{\nu} \pi^{(0)}_A (q^2)
\end{eqnarray}

\noindent we can prove the dispersion relations

\begin{equation}
\pi_V (q^2) - \pi_A (q^2) = {F^2_{\pi} \over q^2} + \int^{\infty}_{4 
m^2_{\pi}} ds' {\rho_V (s') - \rho_A (s') \over s'
- q^2 - i \epsilon}
\end{equation}

\noindent with the imaginary parts conventionally named via

\begin{equation}
\rho_{V/A} (s) = {1 \over \pi} Im 
\pi_{V/A} (s)
\end{equation}

What is known theoretically about these amplitudes?  At low energy, chiral 
perturbation theory predicts the form[4]

\begin{eqnarray}
\pi^{\mu \nu}_V (q^2) - \pi^{\mu \nu}_A (q) &=& \left[ {q_{\mu} q_{\nu}
\over q^2 - m^2_{\pi} + i \epsilon} - g_{\mu \nu} \right]  F^2_{\pi} \\
\nonumber
&+& (q_{\mu} q_{\nu} - g_{\mu \nu} q^2) \left[ {1 \over 48 \pi^2} \left( 1
- {4m^2_{\pi} \over q^2} \right) H (q^2) - 4L^r_{10} \right. \\ \nonumber
 &-& \left.{1 \over 48 \pi^2}  \left( ln {m^2_{\pi}
\over \mu^2} + {1 \over 3} \right) \right] \\ \nonumber 
\rho_V (s) &=& {1 \over 48 \pi^2} \left[ 1 - {4 m^2_{\pi} \over s}
\right]^{3 \over 2} \theta (s - 4 m^2_{\pi}) + O(s) \\ \nonumber
\rho_A (s) &=& {s \over 96 (4 \pi F_{\pi})^2} + O(s^2)
\end{eqnarray}

\noindent Here $L^r_{10}$ is a low energy constant measured in radiative 
pion decay, $\pi \rightarrow e \nu \gamma$.

At high energy, perturbative QCD may be used to 
analyze the two point function.  In 
the chiral limit, $m_q = 0$, which will be used for the rest of this section, 
the operator product expansion can be used to show that the difference 
$\pi_V - \pi_A$ falls as $1 \over q^6$ and $\rho_V (s) - \rho_A (s) \sim 
{1 \over s^3}$.  In terms of four quark operators, which are here evaluated 
in the vacuum saturation approximation[11], one has

\begin{eqnarray}
\pi_V (q^2) - \pi_A (q^2) = {32 \pi \over 9} {\langle \sqrt{\alpha_s} 
\bar{q} q \rangle^2_0 \over q^6} \left\{ 1 + {\alpha_s (q^2) \over 4 \pi} 
\left[ {247 \over 12} + ln {\mu^2 \over -q^2} \right] \right\} \\ \nonumber
\rho_V (s) \rightarrow \rho_A (s) \rightarrow {1 \over 8 \pi^2} \left[ 1 + 
{\alpha_s (s) \over \pi} \right] \; , \; s \rightarrow \infty \\ \nonumber
\rho_V (s) - \rho_A (s) \sim {8 \over 9} {\alpha_s \langle \sqrt{\alpha_s} 
\bar{q} q \rangle^2_{\infty} \over s^3}
\end{eqnarray}

\noindent We see that $\pi_V - \pi_A$ and $\rho_V - \rho_A$ are very well 
behaved at large $q^2, s$.

We can combine up this information to get a set of sum rules.  The 
requirement that, as $q^2 \rightarrow \infty$, there is no ${1 \over q^2}$ 
term in the dispersion relation Eq.24 , requires

\begin{equation}
F^2_{\pi} = \int^{\infty}_0 ds ( \rho_V (s) - \rho_A (s))
\end{equation}

\noindent while the absence of ${1 \over q^4}$ implies

\begin{equation}
0 = \int^{\infty}_0 ds s ( \rho_V (s) - \rho_A (s))
\end{equation}

\noindent These are the Weinberg sum rules[12], the second of which is only 
true in the $m_q \rightarrow 0$ limit.  At low energy, expansion of the 
dispersion integral and chiral results in powers of $q^2$ imply[13,4]

\begin{equation}
-4 \bar{L}_{10} = \int^{\infty}_{4m^2_{\pi}} {ds \over s} ( \rho_V (s) - 
\rho_A (s))
\end{equation}

\noindent with

\begin{eqnarray}
\bar{L}_{10} &=& L^r_{10} (\mu) + {1 \over 192 \pi^2} \left[ ln 
{m^2_{\pi} \over \mu^2} + 1 \right] \\ \nonumber
&=& (-0.7 \pm 0.03) \times 10^{-2} \; (Expt: \pi \rightarrow e \nu \gamma)
\end{eqnarray}

\noindent Here I have given the sum rule for finite $m^2_{\pi}$ since there 
is a behavior proportional to $ln m^2_{\pi}$ at the low energy 
end of the integral.  These sum rules illustrate one of the uses of chiral 
dispersion relations, which is the prediction/calculation of low energy 
constants (here $F_{\pi}$ and $L_{10}$). When these constraints are
satisfied we have an accurate matching of the two descriptions, valid
to $O(E^4)$.

Another use of chiral dispersion relations is in extending the reach of 
calculations and even opening up the possibility of entirely new types of 
calculations.  Consider the Compton amplitude $\gamma \pi \rightarrow 
\gamma \pi$.  In the soft pion limit, chiral symmetry relates this to the 
vacuum polarization tensors

\begin{eqnarray}
&&\lim_{p \rightarrow 0} \langle \pi^+ (p) \mid T (V^{\mu} (x) V^{\nu} 
(0)) \mid \pi^+ (p) \rangle \\ \nonumber
&=& - {1 \over F^2_{\pi}} \langle 0 \mid T (V^{\mu} (x) V^{\nu} (0) - 
A^{\mu} (x) A^{\nu} (0) ) \mid 0 \rangle \\ \nonumber
&=& - {1 \over F^2_{\pi}} \left[ \pi^{\mu \nu}_V (x) - \pi^{\mu \nu}_A (x) 
\right]
\end{eqnarray}

\noindent If one takes the Compton amplitude and ties together the two 
electromagnetic currents with a photon propagator, one obtains the pion 
electromagnetic mass shift.[14]  
Clearly the chiral representation, Eq. 26, would 
be inadequate to calculate this, as the photon loop integral goes over all 
values of $q^2$.  However, after some algebra plus the application
of the Weinberg sum rules, the dispersive representation allows one to 
write this as [14]

\begin{equation}
m^2_{\pi^+} - m^2_{\pi^0} = -{3e^2 \over 16 \pi^2 F^2_{\pi}} 
\int^{\infty}_{0} ds s ln s \left[ \rho_V (s) - \rho_A (s) \right]
\end{equation}

\noindent which is an exact relation in the chiral limit. Note that here 
chiral symmetry was used to relate different amplitudes in Eq. 32
 and to provide
low energy constraints, as in the Weinberg sum rules, while
dispersion relations were needed to provide a predictive
framework for the intermediate energy region.

In a similar way, one can calculate reliably a new weak nonleptonic matrix 
element.[15]  Consider the hypothetical weak Hamiltonian

\begin{equation}
H_V = {g^2_2 \over 8} \int d^4 x i D^{\mu \nu}_F (x, M_w) T \left( 
\bar{d} (x) \gamma_{\mu} u (x) \bar{u} (0) \gamma_{\nu} s (0) \right)
\end{equation}

\noindent Up to some KM factors, this would be the usual weak 
Hamiltonian if the vector currents were replaced by $\gamma_{\mu} (1 + 
\gamma_5)$.  In the chiral limit, we have another chiral sum rule

\begin{equation}
\langle \pi^- \mid H_V \mid K^- \rangle = {3i G_F \over 32 \pi^2 \sqrt{2} 
F^2_{\pi}}A
\end{equation}

\noindent with

\begin{equation}
A = M^2_w \int^{\infty}_0 ds {s^2 ln (s/M^2_w) \over s - M^2_w + i 
\epsilon} \left[ \rho_V (s) - \rho_A (s) \right]
\end{equation}

\noindent which is exact in the chiral limit.

Gene Golowich and I have provided a phenomenological analysis 
of these sum rules.[16]  The physics of the spectral functions $\rho_{V,A}$ is 
basically simple.  At intermediate energies they are measured in $\tau$ decay 
and $e^+ e^-$ annihilation, and the largest features are the $\rho$ and 
$a_1$ resonances, with very much smaller $4 \pi, 5 \pi$ etc. contributions.  
At low energy this can be merged smoothly to chiral predictions and at high 
energy $\rho_V - \rho_A$ vanishes rapidly and we matched the data to QCD 
around $s = 5 GeV^2$.  There are some experimental uncertainties, but 
these can in principle be reduced in the future.

The $L_{10}$ sum rule works well with very little uncertainty as it is 
sensitive to the lowest energy contributions.  The Weinberg sum rules and 
that for $\Delta m^2_{\pi}$ work within the experimental uncertainties.  We 
have proceeded by imposing them exactly on our $\rho_V - \rho_A$, which 
requires only minor adjustments within the allowed error bars.  That this is 
possible is a nontrivial test of the theoretical framework.  Finally the weak 
matrix element is predicted ($A = -0.062 \pm 0.017 GeV^2$).  This can 
perhaps be compared with future lattice calculations.

\section{The Elastic Approximation and the Omnes Problem}

The pion formfactor and the Weinberg sum rules are particularly powerful
because we have a direct measurement of the required imaginary parts.
In many other cases, we do not have this luxury. Nevertheless, much of the
dynamics of the intermediate states follows from the behavior of 
$\pi\pi$ scattering, about which we know a good deal. This allows us to
predict the behavior of the imaginary part of the desired amplitude, with
the modest additional assumption that the only important intermediate state
is the $\pi\pi$ channel, i.e. the elastic approximation.

Consider some two particle amplitude $f(s)$ of a given isospin and angular 
momentum which is analytic in complex $s$ plane except for a cut above 
two particle threshold $s_0 = 4m^2$.  The inelastic thresholds are somewhat 
higher, for example $s_{inel} = 16 m^2$.  In the elastic region, Watson's 
theorem tells us that the phase of the amplitude is that of the corresponding 
two particle scattering amplitude

\begin{equation}
f(s) = e^{i \delta (s)} \mid f(s) \mid
\end{equation}

\noindent In practice inelasticities do not play a significant role in low 
energy pion physics up to 1 GeV ($K \bar{K}$ threshold), and one often 
assumes an approximation of keeping only the elastic channel.  While 
probably reasonable, it is important to realize that the elastic approximation 
relies on more than just Watson's theorem and produces more than just the 
phase of the amplitude.

The Omnes problem[17] is the mathematical exercise of finding functions which 
are analytic except for a cut $4m^2 < s < \infty$, which are real when 
$s$ is real and less than $4m^2$ and for which $f(s) e^{-i \delta (s)}$ is real 
when $s$ is real and greater than $4m^2$.  The solution is given by

\begin{eqnarray}
f(s) &=& P(s) D^{-1} (s) \\ \nonumber
D^{-1} (s) & \equiv & exp \left\{ {s \over \pi} \int^{\infty}_{4m^2} {dt 
\over t} {\delta (t) \over t - s - i \epsilon} \right\}
\end{eqnarray}

\noindent as long as

\begin{equation}
\lim_{s \rightarrow \infty} \delta (s) = finite \; ; \lim_{s \rightarrow \infty} 
{\mid f (s) \mid \over s} \rightarrow 0
\end{equation}

\noindent In the above $P(s)$ is a polynomial in $s$, and $D^{-1} (s)$ is 
called the Omnes function.

Note that this is not exactly the right problem for QCD.  The assumption 
that $f(s) e^{-i \delta (s)}$ is real above the cut implies that the reaction
is elastic at {\em all} energies.  Once inelastic channels open up, the 
quantity $f(s)e^{-i\delta (s)}$ rapidly deviates from being real.
In QCD, once one is above 1 
GeV, the inelastic channels open rapidly and become quite numerous, leading
to perturbative QCD behavior at precociously low energies.  It is not
known how to provide a general solution to the QCD type problem (although the
form of the solution to the two channel problem is also known), nor is it known
how much of an effect the multiple inelasticities of QCD have on the Omnes
function.

\section{Example: $\gamma \gamma \rightarrow \pi \pi $}

The reaction $\gamma \gamma \rightarrow \pi^+ \pi^-$ and $\gamma 
\gamma \rightarrow \pi^0 \pi^0$  are of interest in the development of chiral 
theory because $\gamma \gamma \rightarrow \pi^0 \pi^0$ first arises as a 
pure loop effect as there are not tree level contributions at $O(E^2)$ or 
$O(E^4)$.  For these 
reactions, we have both a one-loop [18]and two loop [19] chiral 
analysis as well 
as dispersive treatments[20,8] and experimental data.  This makes 
these reactions 
excellent illustrations of chiral techniques and of the ties with 
dispersion relations.

The $\gamma \gamma \rightarrow \pi \pi$ matrix elements can be 
decomposed into isospin amplitudes

\begin{eqnarray}
f^{+-}(s) &=& {1 \over 3} [2f_0(s) + f_2(s)] \\ \nonumber
f^{00}(s) &=& {2 \over 3} [f_0(s) - f_2(s)]
\end{eqnarray}

\noindent The dominant partial waves at low energy are the S waves and 
these are predicted in a one loop chiral analysis to be

\begin{eqnarray}
f^{chiral}_I (s) &=& {1-\beta^2 \over 2 \beta} ln \left( {1 + \beta \over 1 - 
\beta} \right) - 
{(1 - \beta^2) \over 4 \pi} t^{CA}_I (s) ln^2 {\beta + 1 \over 
\beta - 1} \\ \nonumber
&& - {1 \over \pi} t^{CA}_I (s) + {2 \over F^2_{\pi}} \left( L^r_9 + 
L^r_{10} \right) s
\end{eqnarray}

\noindent where

\begin{equation}
\beta = \sqrt{1 - {4m^2_{\pi} \over s}}
\end{equation}

\noindent and $t^{CA}_I (s)$ are the lowest order $\pi \pi$ scattering 
amplitudes

\begin{equation}
t^{CA}_0 = {2s-m_{\pi} \over 32 \pi F^2_{\pi}} \; ; t^{CA}_2 = -{(s - 
sm^2_{\pi}) \over 32 \pi F^2_{\pi}}
\end{equation}

The dispersion relation has been derived by Morgan and Pennington[20], in 
terms of an amplitude $p_I (s)$ which has the same left-hand singularity 
structure as $f_I(s)$ but which is real for $s > 0$.  Then $[f_I(s) - p_I(s)] 
D_I(s)$ satisfies a twice subtracted dispersion relation and we have

\begin{equation}
f_I(s) = D^{-1}_I (s) \left[ (c_I + d_I s) + p_I (s) D_I (s) - {s^2 \over \pi} 
\int^{\infty}_{4m^2_{\pi}} {ds' \over s'^2} {p_I (s') 
Im D_I (s') \over s' - s - i \epsilon} \right]
\end{equation}

\noindent with two subtaction constants per channel $c_I, d_I$.  As a 
prelude to the matching we note that Low's theorem requires that $f_I(s)$ 
be the Born scattering amplitude at low energies.  Therefore

\begin{equation}
p_I(s) = f^{Born}_I(s) + O(s) = {1- \beta^2 \over 2 \beta} ln \left( {1 + 
\beta \over 1 - \beta} \right) + O(s)
\end{equation}

\noindent This is the $O(E^2)$ result.  To proceed to order $E^4$ we note 
that the leading piece of $Im D_I(s)$ is also known, i.e.,

\begin{equation}
Im D_I(s) = - \beta t^{CA}_I (s)
\end{equation}

\noindent as this is the lowest order $\pi \pi$ scattering amplitude.  Using 
this, the dispersive integral can be done exactly, leading to

\begin{eqnarray}
f_I(s) &=& D^{-1} (s) \left[ c_I + s \left( d_I - {t^{CA}_I (0)
\over 12 \pi  m^2_{\pi}} \right) + D_I(s) {1 - \beta^2 \over 2 \beta}
ln \left( {1 + \beta  \over 1 - \beta} \right) \right. \\ \nonumber
&-&\left. {1 \over 4 \pi } (1 - \beta^2) t^{CA}_I (s) ln^2 \left( {\beta
+ 1 \over \beta - 1} \right) \right] + \ldots
\end{eqnarray}

\noindent A comparison of this with the $O(E^4)$ chiral results then 
indicates that this procedure has reproduced all of the one loop results, as 
long as we choose the subtraction constants as[8]

\begin{equation}
c_I = 0 \; ; d_I = {2 \over F^2_{\pi}} \left( L^r_9 + L^r_{10} \right) + 
{t^{CA}_I (0) \over 12 \pi m^2_{\pi}}
\end{equation}

\noindent Again we see that the dynamical content of the one loop chiral 
calculation is also contained in the dispersive treatment when the imaginary 
part is taken to be the lowest order scattering amplitude.  However, chiral 
symmetry also predicts the subtraction constants, which in this case are 
known from measurements in radiative pion decay.

Having identified the subtraction constants one can add the ingredients to 
complete the calculation.  The most important at threshold is the use of the 
real world $D^{-1}_I (s)$[7].  This change alone produces a significant effect 
in the amplitude 
even near threshold in the neutral case.  
The second step is a better determination of 
$p_I(s)$ which includes the $O(E^4)$ chiral corrections to it as well as the 
$\rho, \omega, A1$ poles which are known (from $\rho \rightarrow \pi 
\gamma$ etc. data) to occur in the Compton amplitude. 
Figure 3 shows the data for the 
reaction $\gamma \gamma \rightarrow \pi^0 \pi^0$ along with the one-loop 
chiral prediction (dashed line) and the modification obtained by the dispersive
treatment (solid line). The one-loop 
chiral result is of the right rough size, its slope is low at 
threshold and it grows unphysically at high energy.
Near threshold  the difference in the two calculations
comes almost exclusively from the rescattering corrections generated through
the dispersion relation. The change is sizable even at low energy, since the
rescattering in in the $I=0, J=0$ channel. The Omnes function 
alone has brought the threshold region into better agreement with the data.
It has also  tamed the 
high energy growth. The final result 
(with no free parameters) matches the data very well, and also 
gives the charged channel correctly.

Bellucci, Gasser and Sainio[19] have performed the enormously difficult two 
loop calculation.  [In fact, technically they employ dispersive methods to do 
portions of this.]  At two loop order, new low energy constants appear, 
which are not measured in any other process.  Therefore the authors have to 
step outside of pure chiral perturbation theory in order to model these 
constants, using vector meson dominance.  Much like the dispersive work 
described above, these constants play little role in the threshold region, but 
are important for the shape of the amplitude for moderate energy.  It is 
very interesting that their results look very similar to the 
dispersive treatment described above.  

Both of these methods have potential limitations.  In principle, the only 
limitation of the dispersive treatment is the fact that it can miss $O(E^6)$ 
terms in the subtraction constants $c_I, d_I$.  These would be corrections 
to results given above by factors of $m^2_{\pi}$ or $m^2_{\pi} ln 
m^2_{\pi}$.  In practice we also need to model the higher order terms in 
$p_I(s)$.  As for the two loop chiral result, its 
only limitation in principle is 
the fact that it misses higher order dependence in the energy.  By 
construction it is valid to order $E^6$, but does not contain 
higher order $s$ dependence, and 
so would be expected to fall apart soon after the $E^6$ dependence became 
important.  In practice, this approach also needs to do some modeling in 
order to estimate the unknown low energy constants.  The fact that the 
results agree so well with each other and with the data indicates that these 
limitations are not very important at these energies.  Both capture the 
important physics, and do so in a reasonably controlled fashion.  There is of 
course a significant practical advantage to the dispersive approach--it is far 
easier!

\section{Other uses of dispersion relations in effective field
theories}

The examples given above have all been in the context of chiral
perturbation theory. We have a detailed knowledge of the behavior of
chiral amplitudes at low energy as well as measurements of various
reactions which can be used in dispersive integrals. However, other
effective field theories can also benefit from the dispersive method
of analysis.
 
The main obstacle to many applications of dispersion relations is the
lack of an experimental measurement of the imaginary part of the 
amplitude. This will of course preclude a rigorous model independent
calculation, but one may still be able to use approximations or
modeling to make progress on the problem. Dispersive calculations
open up new methods of approximation.
 
An example is in the electroweak corrections due to TeV scale physics.
Peshkin and Takeuchi have defined a variable $S$ which describes some
of the low energy effects on $W$ and $Z$ propagation due to new
physics, and have given a dispersive sum rule for it[21]. The sum rule
is equivalent to the $L_{10}$ sum rule, Eq. 30, where the vacuum
polarization functions are those for the $W$ and $Z$ self-energies.
These sum rules have been applied extensively to Technicolor theories.[22]
One potential new application is to the effect of a heavy Higgs boson.
While this has been calculated in detail in perturbation theory[23], as
the  Higgs mass gets heavier, perturbation theory ultimately becomes
inapplicable because the theory becomes strongly coupled. However in
the strongly coupled regime we still know various features of the full
amplitudes, such as threshold behavior and unitarity bounds, even if
we cannot calculate precisely. Preliminary analysis suggests that that
one can use the dispersive sum rule plus the general behavior of the
amplitude to argue that the perturbative estimate is not far wrong[24].

There have also been a set of interesting applications of dispersive
techniques to the Heavy Quark Effective Theory[25]. It is not possible
to summarize all the types of applications that dispersion relations
have in effective field theory. However, the main point of this
lecture is that these techniques are well adapted to the type of
questions that we ask in effective field theories and can sometimes be
useful in extracting features that are not visible in the usual
Feynman diagram methods.

\section{Summary}

We have seen how dispersion relations can add power to effective field
theory.  At its best it uses more physics input.  It can match all 
perturbative effects 
to whichever order that they are known, and can be used to replace the 
modeling of unknown physics by using data instead of models.  However 
there are some limitations, coming both from incomplete data and from the 
fact that we can only determine the subtraction constants to a given order in 
the energy expansion.

The technology for combining these techniques is now developed.  This 
involves first knowing the chiral analysis of the amplitude to a given order 
in the energy expansion.  One also needs a dispersion relation for the 
amplitude in question.  The number of subtractions is determined partially 
by the high energy behavior of the amplitude, but the use of more subtractions 
than are required can help in the matching with the effective field
theory chiral result.  The 
matching occurs order by order in the energy expansion.  When it can be 
done, it is preferable to perform the matching at $O(E^4)$ 
because the resulting dispersive treatment 
is less sensitive to what happens at high energy since a twice subtracted 
dispersion relation can be used.  Finally, the real world data has to be found 
to use in the dispersive integral.  Often, in 
chiral perturbation theory, the use of the elastic approximation 
is made for this, allowing the use of known $\pi \pi$ scattering data.

The output of these efforts can be several.  Most commonly, these 
techniques are used to extend the range and accuracy of the chiral 
calculations, by getting around the limitation of the energy expansion.  The 
method can be used to predict unknown chiral coefficients, as was shown 
for the case of $L_{10}$.  We can use these techniques to remove or reduce 
the model dependence of some result.  Finally, dispersive techniques allow 
us to perform completely new types of calculations, such as the hadronic 
matrix elements of Section 5.

{\bf Acknowledgements}.  I would like to thank Juerg Gasser, Barry 
Holstein and Heiri Leutwyler for sharing their insights into dispersion
relations and chiral symmetry.  In addition, I thank the organizers,
lecturers and students for making this school so enjoyable and
productive.

{\bf References}.\\
1) An earlier version of the present lecture has appeared in {\it Chiral
Dynamics of Hadrons and Nuclei}, ed by D.P. Min and M. Rho (Seoul
National University Press, Seoul, 1995), p87 (hep-ph/9506205).\\
2) R. Kronig, J. Op. Soc. Am. {\bf 12}, 547 (1926); H. A. Kramers,
Atti Cong. Int. Fisici Como (1927).\\
G. Barton, {\it Dispersion Techniques in Field Theory}, (Benjamin, NY, 
1965).\\
3) J. F. Donoghue, E. Golowich and B. R. Holstein, {\it Dynamics of the
Standard Model}, (Cambridge University Press, Cambridge, 1992).\\
4) J. Gasser and H. Leutwyler, Nucl. Phys. {\bf B250}, 465(1985).\\
5) T. N. Truong, Phys. Rev. Lett. {\bf 61},2526 (1988).\\
6) As far as I know, the first attempt to use dispersion relations to
improve the predictions of chiral perturbation theory occurred in Ref. 7.\\
7) J. F. Donoghue, J. Gasser and H. Leutwyler, Nucl. Phys. {\bf B343},
341(1990).\\
8) J. F. Donoghue and B. R. Holstein, Phys. Rev. {\bf D48}, 137 (1993).\\
9) J. F. Donoghue, C. Ramirez and G. Valencia, Phys.Rev.{\bf D39},1947,1989. \\
10) G. Ecker, J. Gasser, A. Pich, E. de Rafael, Nucl.Phys. {\bf B321},
311,1989. \\
11) M. A. Shifman, A. I. Vainshtein, 
and V. I. Zakharov, Nucl. Phys. {\bf B147},
385 (1979).\\
L. V. Lanin, V. P. Spiridonov, and K. G. Chetyrkin, Sov. J. Nucl. Phys.
{\bf 44}, 892 (1986).\\
12) S. Weinberg, Phys. Rev. Lett. {\bf 17}, 616 (1966).\\
13) T. Das, V. Mathur, and S. Okubo, Phys. Rev. Lett. {\bf 19}, 859
(1967).\\
14) T. Das, G. S. Guralnik, V. S. Mathur, F. E. Low, and J. E. Young,
Phys. Rev. Lett. {\bf 18}, 759 (1967).\\
15) J. F. Donoghue and E. Golowich, Phys. Lett. {\bf 315}, 406 (1993).\\
16) J. F. Donoghue and E. Golowich, Phys. Rev. {\bf D49}, 1513 (1994).\\
17) R. Omnes, Nuovo Cim. {\bf 8}, 1244(1958).\\
18) J. Bijnens and F. Cornet, Nucl. Phys. {\bf B296}, 557 (1988).\\
J. F. Donoghue, B. R. Holstein and Y. C. Lin, Phys. Rev. {\bf D37}, 2423 
(1988).\\
19) S. Bellucci, J. Gasser and M. Sainio, Nucl. Phys. {\bf B423},80{1994).\\
20) D. Morgan and M. R. Pennington, Phys. Lett. {\bf B272}, 134 (1991).\\ 
21) M. E. Peskin and T. Takeuchi, Phys.Rev.Lett.{\bf 65}, 964(1990.) 
22) M. E. Peskin and T. Takeuchi, Phys.Rev.{\bf D46}, 381(1992). 
B. Holdom, J. Terning, Phys.Lett. {\bf B247}, 88(1990). \\
23)  M. J. Herrero, E. Ruiz Morales, Nucl.Phys.{\bf B437}, 319(1995). \\
M. J. Herrero, E. Ruiz Morales, Nucl.Phys.{\bf B418}, 431(1994). \\
A. C. Longhitano, Phys.Rev.{\bf D22}, 1166(1980). \\
T. Appelquist, C. Bernard, Phys.Rev.{\bf D22}, 200(1980). 
\\
24) J. F. Donoghue, work in preparation\\
25) C. G. Boyd, B. Grinstein, R. F. Lebed, UCSD-PTH-95-15, hep-ph/9508242,\\ 
C. G. Boyd, B. Grinstein, R. F. Lebed,
Phys.Lett.{\bf B353}, 306-312 (1995). 
\\
\end{document}